\documentstyle[12pt,psfig,aaspp4]{article}
\def\degr{\hbox{$^\circ$}}
\def\arcmin{\hbox{$^\prime$}}

\def\fs{\hbox{$.\!\!^{\rm s}$}}

\def\farcm{\hbox{$.\mkern-4mu^\prime$}}
\def\farcs{\hbox{$.\!\!^{\prime\prime}$}}         
\def\gsim{\mathrel{\hbox{\rlap{\lower.55ex \hbox {$\sim$}}
                   \kern-.3em \raise.4ex \hbox{$>$}}}}
\def\lsim{\mathrel{\hbox{\rlap{\lower.55ex \hbox {$\sim$}}
                   \kern-.3em \raise.4ex \hbox{$<$}}}}
\newcommand{\gpm}[3]{$#1^{+#2}_{-#3}$}
\begin{document}  
\title{The Rapid Decay of the Optical Emission from GRB\,980326 and
its Possible Implications}
\author{P.J. Groot\altaffilmark{1}, 
T.J. Galama\altaffilmark{1},        
P.M. Vreeswijk\altaffilmark{1}, 
R.A.M.J. Wijers\altaffilmark{2}, 
E. Pian\altaffilmark{3}, 
E. Palazzi\altaffilmark{3},
J. van Paradijs\altaffilmark{1,4}, 
C. Kouveliotou\altaffilmark{5,6},
J.J.M. in 't Zand\altaffilmark{7},
J. Heise\altaffilmark{7},
C. Robinson\altaffilmark{4,6}
N. Tanvir\altaffilmark{2},
C. Lidman\altaffilmark{8},
C. Tinney\altaffilmark{9},
M. Keane\altaffilmark{10},
M. Briggs\altaffilmark{4,6},
K. Hurley\altaffilmark{11},
J.-F. Gonzalez\altaffilmark{8},
P. Hall\altaffilmark{12},
M.G. Smith\altaffilmark{10},
R. Covarrubias\altaffilmark{10},
P. Jonker\altaffilmark{1},
J. Casares\altaffilmark{13},
F. Frontera\altaffilmark{3},
M. Feroci\altaffilmark{14},
L. Piro\altaffilmark{3},
E. Costa\altaffilmark{3},
R. Smith\altaffilmark{15},
B. Jones\altaffilmark{16},
D. Windridge\altaffilmark{16},
J. Bland-Hawthorn\altaffilmark{9},
S. Veilleux\altaffilmark{17},
M. Garcia\altaffilmark{18},
W.R. Brown\altaffilmark{18},
K.Z. Stanek\altaffilmark{18},
A.J. Castro-Tirado\altaffilmark{19,20},
J. Gorosabel\altaffilmark{19},
J. Greiner\altaffilmark{21},
K. J\"ager\altaffilmark{22},
A. B\"ohm\altaffilmark{22},
K.J. Fricke\altaffilmark{22}
}

\altaffiltext{1}{Astronomical Institute `Anton Pannekoek', University
of Amsterdam,
\& Center for High Energy Astrophysics,
Kruislaan 403, 1098 SJ Amsterdam, The Netherlands}
\altaffiltext{2}{Institute of Astronomy, Madingley Road, Cambridge CB3 0HA, UK}
\altaffiltext{3}{CNR Bologna, Via P. Gobetti 101, 40129 Bologna, Italy}
\altaffiltext{4}{Physics Department, University of Alabama in
Huntsville, Huntsville AL 35899, USA}
\altaffiltext{5}{Universities Space Research Association}
\altaffiltext{6}{NASA/MSFC, Code ES-84, Huntsville AL 35812, USA}
\altaffiltext{7}{Space Research Organisation of the Netherlands
(SRON), Sorbonnelaan 2, Utrecht, The Netherlands }
\altaffiltext{8}{ESO, Casilla 19001, Santiago 19, Chile}
\altaffiltext{9}{Anglo-Australian Observatory, PO Box 296 Epping, NSW
2121, Australia} 
\altaffiltext{10}{Cerro Tololo Interamerican Observatory, Casilla 603,
La Serena, Chile}
\altaffiltext{11}{UC Berkeley, Space Sciences Laboratory, Berkeley, 
CA 94720-7450, USA}
\altaffiltext{12}{Department of Astronomy, University of Toronto, 60
St. George Street, Toronto, Ontario M5S 3H8, Canada}
\altaffiltext{13}{Instituto Astrof\'{\i}sica de Canarias, Tenerife,
Spain}
\altaffiltext{14}{Istituto di Astrofisica Spaziale, CNR, Via Fosso del
Cavaliere, Roma, I-00133, Italy}
\altaffiltext{15}{University of Wales, Cardiff, UK}
\altaffiltext{16}{University of Bristol, Bristol, UK}
\altaffiltext{17}{Department of Astronomy, University of Maryland,
College Park, MD 20742, USA}
\altaffiltext{18}{Center for Astrophysics, 60 Garden St., Cambridge, MA
02138, USA}
\altaffiltext{19}{Laboratorio de Astrof{\'\i}sica Espacial y
F{\'\i}sica Fundamental
(LAEFF-INTA), P.O. Box 50727, E-28080 Madrid, Spain}
\altaffiltext{20}{Instituto de Astrof{\'\i}sica de Andaluc{\'\i}a
(IAA-CSIC), 
P.O. Box 03004, E-18080 Granada, Spain}
\altaffiltext{21}{Astrophysikalisches Institut Potsdam, D-14482
Potsdam, Germany}
\altaffiltext{22}{Universit\"ats-Sternwarte G\"ottingen, Geismarlandstr. 11, 
D-37083 G\"ottingen, Germany}

\begin{abstract}
We report the discovery of the optical counterpart to GRB\,980326. Its
rapid optical decay can be characterized by a power
law with exponent --2.10$\pm$0.13 and a constant underlying source at 
$R_{\rm c}$=25.5 $\pm$0.5. Its optical colours 2.1 days
after the burst imply a spectral slope of --0.66$\pm$0.70.
The $\gamma$-ray spectrum as observed with BATSE shows that it is among
the 4\% softest bursts ever recorded. We argue that the rapid optical
decay may be a reason for the non-detection of some low-energy
afterglows of GRBs. 
\end{abstract}


\setcounter{footnote}{0}
\section{Introduction}
The redshift determinations for
GRB\,970508 (Metzger et al., 1997) and GRB\,971214 (Kulkarni et al., 1998)
have demonstrated that GRBs originate at 
cosmological distances and are therefore the most
powerful photon sources in the Universe, 
with peak luminosities exceeding 10$^{52}$ erg/s, assuming isotropic emission. 
Afterglow studies of GRB\,970228 (Galama et
al., 1997, 1998a), GRB\,970508 (Galama et al., 1998b, c, d; Pedersen
et al., 1998; Castro-Tirado et al., 1998a), and GRB\,971214 
(Halpern et al., 1998, Diercks et al., 1998) show a generally good
agreement with fireball model predictions 
(Wijers, Rees and M\'eszar\'os, 1997; Sari, Piran and Narayan, 1998,
hereafter SPN98). 

There are, however, a few marked cases where no X-ray or optical
afterglow is seen, most notably GRB\,970111 (optical:Castro-Tirado et al.,
1997; Gorosabel et al., 1998, X-rays, debated: Feroci et al.,1998),
GRB\,970828 (optical: Groot et al., 1998a) and GRB\,980302 (X-rays). 
In the last case, RXTE/PCA scanning,
starting only 1.1 hours after the burst, found no X-ray afterglow 
at a level $>$1 mCrab. 
One possible explanation for the lack of optical counterparts is the 
extinction by large column densities of gas and dust, 
obscuring the GRB afterglows (Groot et al., 1998a;
Halpern et al., 1998). This might indicate an origin in
star-forming regions where large quantities of gas and dust are
present (e.g. Paczy\'nski, 1998). However, this scenario does not so
readily explain the non-detection of an X-ray afterglow.  

GRB\,980326 was detected (Celidonio et al., 1998) on Mar. 26.888 UT with 
one of the Wide Field Cameras (WFCs; Jager et al., 1997) and the 
Gamma Ray Burst Monitor (GRBM; Frontera et al., 1997; Feroci et al.,
1997) on board BeppoSAX (Piro, Scarsi and Butler, 1995), 
with Ulysses (Hurley et al., 1998) and with the 
Burst and Transient Source Experiment (BATSE; Briggs et al., 1998) on
board the Compton Gamma Ray Observatory. Its best WFC position is 
RA= 08$^{\rm h}36^{\rm m}26^{\rm s}$, Decl = --18\degr53\farcm0 
(J2000), with an 8\arcmin\ (radius) accuracy. 
RXTE/PCA scanning 8.5 hours after the burst sets an upper limit of
1.6$\times$10$^{-12}$ erg cm$^{-2}$ s$^{-1}$ on the 2--10 keV X-ray
afterglow of GRB\,980326 (Marshall and Takeshima 1998).
Time-of-arrival analysis between the Ulysses spacecraft,
BeppoSAX and BATSE, allows the construction of an Interplanetary Network (IPN)
annulus which intersects the BeppoSAX WFC camera error box (Hurley et
al., 1998). The combined WFC/IPN  error box is shown in Fig.\ \ref{fig:ipnfc}. 

In the BATSE energy range (25--1800 keV) the event lasted $\sim5$s,
is resolved into three narrow peaks, with a peak flux of 
8.8$\times$10$^{-7}$ ergs cm$^{-2}$ s$^{-1}$, over a 1s timescale. 
This places it at the knee
of the log$N$-log$P$ distribution (Meegan et al. 1996). Its total
25--1800 keV fluence was 1.4$\times10^{-6}$ ergs cm$^{-2}$. 
The event averaged spectrum has a shape typical of
GRBs (photon index \gpm{-3.1}{0.25}{0.5}), 
but its $E_{\rm peak}$, where the $\nu F_{\nu}$
spectrum peaks, is unusually low: $E_{\rm peak}$=47$\pm5$ keV. Only 4\%
of the bursts in the 
sample of Mallozzi et al. (1998, over 1200 GRBs) have smaller
$E_{\rm peak}$ values. However, Mallozzi et al. have also shown that
there is a correlation between GRB intensity and spectral hardness
(expressed in $E_{\rm peak}$ values). 
For bursts with similar peak fluxes,
the smallest $E_{\rm peak}$ value there is $\sim$ 70 keV 
(Mallozzi, private communication), which
demonstrates the exceptional softness of the integrated spectrum of
GRB\,980326.

\section{The optical counterpart}

\begin{figure*}
\caption[]{The combined BeppoSAX WFC and IPN arc error box for
GRB\,980326, an AAT Mar. 27.4 UT,
 1\farcm6$\times$1\farcm6 $R_{\rm c}$-band finding chart of the
field of the optical transient and a small inset of the immediate surroundings of the OT, made from addition of the last three NTT nights. The solid
IPN annulus is the BeppoSAX/Ulysses (S/U) annulus, the dotted annulus
is the BATSE/Ulysses (B/U) annulus.  
Local comparison stars are indicated by no. 1--4.\label{fig:ipnfc}}
\centerline{\psfig{figure=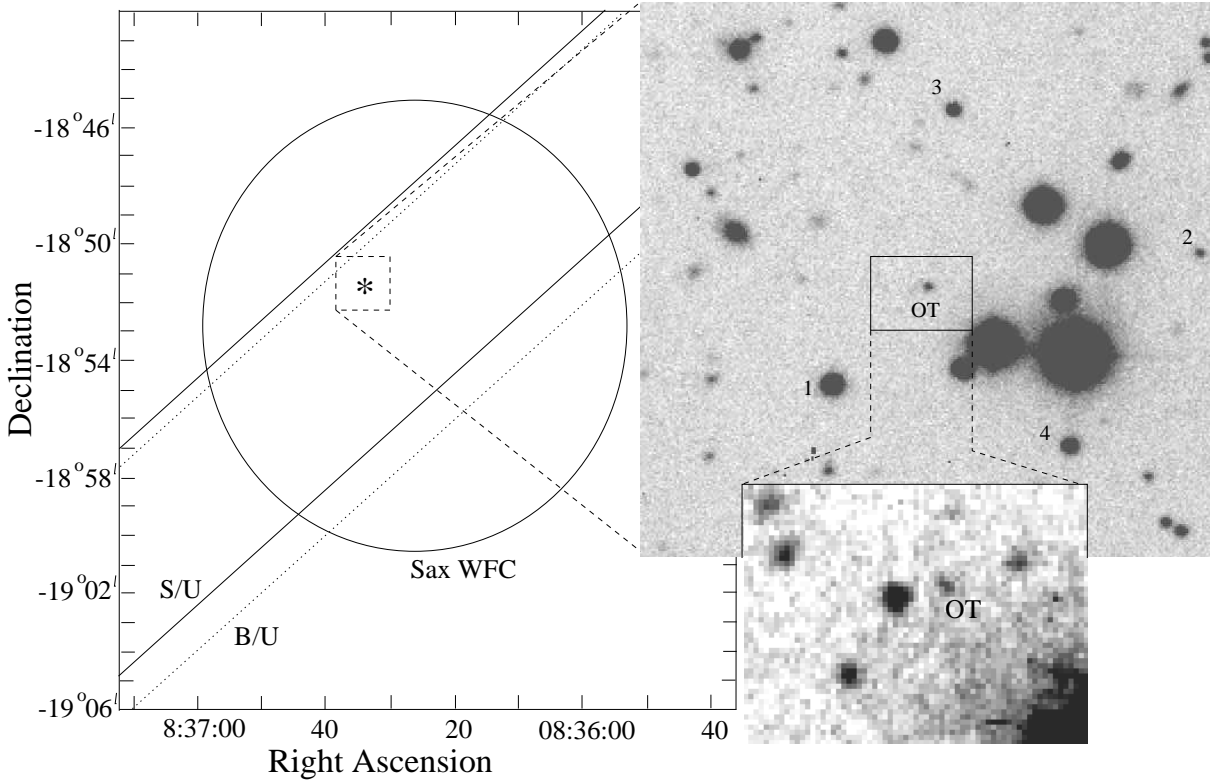,width=16cm,angle=-90}}
\end{figure*}

Optical Cousins $R_{\rm c}$-band observations started at the Anglo-Australian
Telescope (AAT) on Mar. 27.40 UT, followed by observations at
the 3.5m New Technology Telescope (NTT) and the 1.54m Danish telescope
(1.5D) at ESO (Chile), the 4m Victor Blanco telescope at 
CTIO (Chile), the Fred Lawrence Whipple 1.2m
(FLW 1.2m; USA) telescope, the 1.5m Bologna University (BO; Italy) telescope
and the 2.2m
Calar-Alto (CAHA 2.2m; Spain) telescope (see Table 1). 
All observations were debiased and flatfielded in the standard
fashion. Table 2 shows the magnitude of the comparison stars in all photometric
bands used. Note that star 2 (see Fig.\ \ref{fig:ipnfc}) was not
detected in the $B$-band calibration frames. 

From a comparison of the first observations at the AAT
and ESO/CTIO we discovered one clearly variable object (Groot et al., 1998b). 
Its location is RA=08$^{\rm h}36^{\rm m}34\fs28$, 
Dec = --18\degr51\arcmin23\farcs9 (J2000) with an 0\farcs4 accuracy.
Fig.\ \ref{fig:ipnfc} shows the region of the OT. 
Aperture photometry on the combined
WFC/IPN  error box for the first AAT and CTIO epoch found, apart from 
asteroid 1998 FO 126 at $R_{\rm c}$=22.7, no other
object with a change in magnitude $>$0.4 mag down to $R_{\rm c}$=23. 
Although the
variability of sources at $R_{\rm c}>$ 20 is very poorly known, 
we conclude that the optical transient is the counterpart to 
GRB\,980326, also considering the exhibited power law decay.

\begin{figure}
\caption[]{$R_{\rm c}$-band light curve of GRB\,980326. All errors are
1$\sigma$, all upper limits are 3$\sigma$. The dashed line indicates
the power law decay and constant source fit (see Sect. 2).
 \label{fig:light}}
\centerline{\psfig{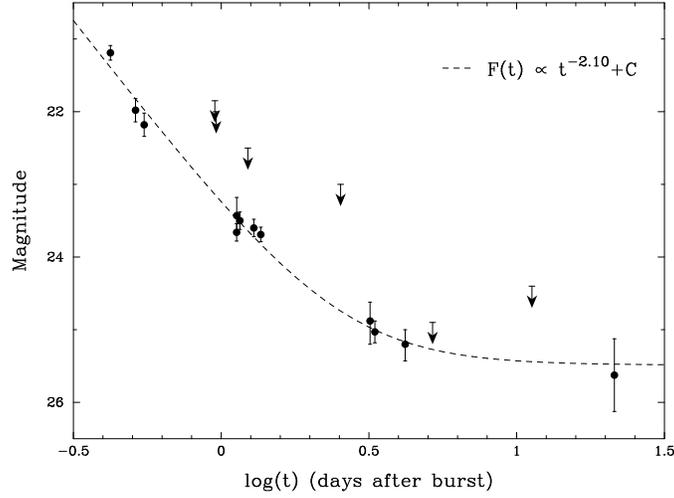}}
\end{figure}

Figure\ \ref{fig:light} shows the $R_{\rm c}$-band light curve of the optical
transient. It exhibits a temporal decay which, as applied in previous
bursts, can be fitted with a power law and a constant 
source: $F_{\nu} \propto t^{-\alpha}$ + C. 
The power law exponent, 
$\alpha$ = 2.10$\pm$0.13, is by far higher than that of previous afterglows.
The light curve exhibits a flattening, with a fitted
constant source of $25.5\pm0.5$ ($\chi^2$ for the fit is 10.2/9), such as
observed for GRB\,970508 (Pedersen et al., 1998; 
Garcia et al., 1998; Castro-Tirado et al., 1998b), 
which is possibly the signature of an underlying host galaxy.  
Grossan et al. (1998) reported an elongation in the NE-SW direction, which 
is also suggested by visual inspection of the NTT observations taken April
1.08 UT, but S/N levels are too low to draw any conclusion. 
Visual inspection of the observations reported by Djorgovski et al.
(1998) displays an elongation in exactly the perpendicular direction
(SE-NW), which may be an effect of fading of the optical transient. 
This would mean that it is not in the center of an underlying galaxy. 

On the night of Mar. 29.0 UT broadband $BVI_{\rm c}$ measurements of
the optical transient were made at the NTT ($V$ and $I_{\rm c}$) 
and at CTIO ($B$). 
From the fit to the light curve presented in Fig.\ \ref{fig:light} we
deduce an $R_{\rm c}$-band value of 24.50 $\pm$ 0.10 at Mar 29.0 UT. 
The colours of the transient at this time were $B-R_{\rm c}$ = 0.53$\pm$0.34,
$V-R_{\rm c}$ $>$ --0.25, $R_{\rm c}-I_{\rm c} < 2.1$ (3$\sigma$ limits on 
$V$ and $I_{\rm c}$). The $B-R_{\rm c}$ value implies an, uncertain,
 spectral power law index, $F(\nu) \propto \nu^{-\beta}$, of 
$\beta=0.66\pm$0.70. One has to realise 
though, that the underlying source might contribute significantly to the 
colours, depending on the difference between the afterglow and constant source
spectrum. 

\section{Constraints on the electron distribution}

Afterglow observations of GRBs over the last year show that a
relativistic blast wave, in which the highly relativistic electrons
radiate via the synchrotron mechanism, 
provides a generally good description of the observed properties
(Wijers, Rees and Mesz\'ar\'os,1997; SPN98). 
Here we will discuss briefly the implications of the power-law decay exponent 
$\alpha$ and the optical spectral slope $\beta$ for a number of different 
blast wave models. For an extensive discussion on blast wave models and their 
application to GRB afterglows we refer the reader to 
Wijers, Rees and Mesz\'ar\'os (1997), SPN98 and Galama et al. (1998c).

All models have that the flux $F(\nu,t)\propto t^{-\alpha}\nu^{-\beta}$ 
for a range of frequencies and times which contain no spectral breaks.
In each model or spectral state of a model $\alpha$ and $\beta$ are 
functions only of $p$, the power law exponent of the electron Lorentz factor 
($\gamma_e$) distribution, $N(\gamma_e) \propto \gamma_e^{-p}$. 
The measurement of either one of $\alpha$ or $\beta$ therefore
fixes $p$, and predicts the other one. 

Given the poor constraint on the spectral slope, we cannot uniquely
fit 
GRB\,980326, but we will examine whether its rapid decay
requires special circumstances. 
First we assume that both the peak frequency $\nu_{\rm m}$ and the cooling 
frequency $\nu_{\rm c}$ 
(see SPN98 for their definitions) have passed the optical passband at 
0.5 days. 
In this case $p=(4\alpha+2)/3=3.5\pm0.1$, and
$\beta$=$p$/2 = 1.75 $\pm$0.06. 
The second possibility is when $\nu_{\rm m}$ has already passed the
optical at 0.5 days, but $\nu_{\rm c}$ not yet at 4.2 days. 
In this state $p = (4\alpha+3)/3 = 3.8\pm0.1$, and $\beta = -(1-p)/2$
= 1.4 $\pm$0.06. Although the latter case agrees slightly better with the
measured $B-R_{\rm c}$ spectral slope, we are hesistant to draw any conclusion
from this, considering the uncertainty of the spectral slope.  
Both, however, imply a much steeper electron spectrum for this burst than 
the value $p=2.2$ derived for GRB\,970508 (Galama et al. 1998c, d).
In case the blast wave is jet-like, the inferred electron spectrum will 
only be different if the opening angle, $\theta$, of the jet is less than 
the inverse of the opening angle, here $<7^\circ$, in which case for 
slowly cooling electrons $p=\alpha=2.1$,
and for rapidly cooling electrons $p=\alpha-1=1.1$ (Rhoads 1998). In both cases
$\beta=0.55\pm0.05$, consistent with the optical colour.
Values of $p$ less than 2 are often considered implausible, because
they imply a very efficient acceleration mechanism in which the most
energetic electrons carry the bulk of the energy. 

\section{The maximum value of $p$}

What is the maximum value of $p$ that can be reached in shock
acceleration? In non-relativistic strong shocks it is generally accepted that
$p\sim$2 (Bell, 1978; Blandford and Ostriker,
1978). In ultra-relativistic shocks however, the
situation is not so clear (Quenby and Lieu, 1989). 
Recent calculations show that in this case $p$ will be between 
3.2 and 3.8, depending on the morphology of the magnetic field
(Achterberg and Gallant, 1998). This is,
however, when the electrons do not radiate an appreciable part of
their energy during shock acceleration. If the electrons do radiate 
significantly, as is suggested GRB\,970508 (Galama et al., 1998c,d;
SPN98), the electron spectrum will
steepen and the distribution of electrons will no longer be a pure
power law. In a power-law model fit, measured values exceeding
$p\sim3.8$ are therefore expected and as a consequence, power
law decays of afterglows that are even more rapid than the 
$\alpha$=2.10 found here are entirely possible.    
  
\section{Explanations for non-detections: rapid decays and 
galactic halos}

The optical behaviour of bursts like GRB\,970828 (Groot et al., 1998a)
 and GRB\,971214 (Halpern et al., 1998) can be explained by extinction 
due to gas and dust between the observer and the origin of the GRB source. 
However extinction will fail to explain the non-existence of an 
X-ray afterglow above 4--5 keV since at these energies extinction
is negligible.
The fact that all BeppoSAX NFI follow-ups have detected an X-ray
afterglow (with the possible exception of GRB\,970111, Feroci et al.,
1998) and that only two RXTE/PCA scannings (for GRB\,970616 and
GRB\,970828) have produced X-ray afterglows, makes the question arise 
what the cause of this difference is. 

Suppose we have an X-ray afterglow that decays as a power law with 
exponent $\alpha$. What is the X-ray afterglow flux needed 
shortly ($\sim$ 1 minute) after the burst, as a function of $\alpha$, 
if we want to detect the afterglow at a level of $\sim$ 1mCrab after a 
few hours? The X-ray flux after 1 minute 
can be estimated by the X-ray emission detected in the burst 
itself, since this X-ray emission will be a mixture of the X-ray tail of the 
GRB and the start of the X-ray afterglow. We can therefore derive 
an estimate of the upper limit to the X-ray afterglow level 
after a few hours from the prompt X-ray emission. 
   
\begin{figure}
\caption[]{The X-ray flux needed after 1 minute to detect a GRB after 1 (solid
line) or 2 (dashed line) and 5 (dashed-dotted line)
hrs at a level of 1 mCrab as a function of
temporal decay power law index $\alpha$. Indicated for several bursts
with measured $\alpha$ is the {\it total} X-ray flux during the GRB
event. References: GRB\,970228 Costa et al., 1997; GRB\,970402 Nicastro et
al., 1997; GRB\,970508 Galama et al., 1998a, Sokolov et al., 1998;
GRB\,970828 Yoshida et al., 1998; GRB\,971214 Halpern et al., 1998,
Diercks et al., 1998; GRB\,980326 this paper; GRB\,980329 In 't Zand et
al., 1998. \label{fig:fx1m}}
\centerline{\psfig{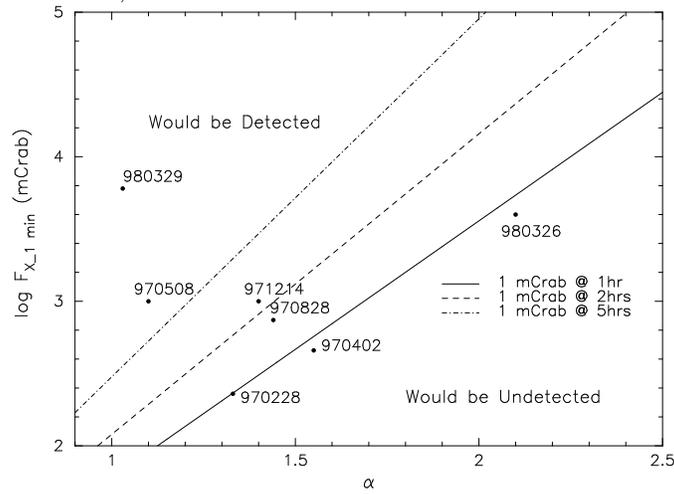}}
\end{figure}

Figure\ \ref{fig:fx1m} shows the flux needed after 1 minute
for a detection after 1, 2 and 5 hours at a level of 1 mCrab as a
function of decay rate $\alpha$. 
For bursts that have detected X-ray or optical
afterglows we have also plotted in Fig.\ \ref{fig:fx1m} the
observed total X-ray fluxes during the bursts versus the X-ray power law
decay index $\alpha$. (For 
GRB\,980326 we used the optical $\alpha$, since no X-ray 
afterglow decay index is known.) 
Because of the mixture explained above these points actually comprise a set of
upper-limits for the flux in the X-ray afterglow after one minute. 
It is not only clear from Fig.\ \ref{fig:fx1m} that most of the bursts
that have been found to exhibit an X-ray afterglow would have been
missed by an RXTE/PCA scan after 2--5 hours, but also that this is
particularly the case for bursts with high values of $\alpha$. 
A rapid decay is therefore a viable explanation for the non-detection
of bursts, even as bright as GRB\,980203, by the current RXTE/PCA
follow-up. 
It has to be noted that the scanning of the RXTE/PCA is
often performed over no more than the 1.5--2$\sigma$ BATSE errorboxes,
and there exists therefore a 5--14\% chance of not scanning the GRB. 

For bursts that show neither X-ray nor optical afterglows,
a different explanation may be found in the fact that all
five detected optical afterglows are associated with galaxies.
In the merging neutron-star scenario, a substantial fraction of bursts
would occur in a galactic halo, where the average density of the 
interstellar medium is $\sim$1000 times less than in a disk. 
Since the afterglow peak flux, 
$F_{\rm m}$, depends on the square root of the density of the ambient 
medium, this would mean a reduction of the afterglow peak flux by
several magnitudes with respect to bursts that go off in higher
density regions (M\'esz\'aros and Rees, 1997).
Since GRBs are detected by their prompt $\gamma$-ray emission, 
probably produced by internal shocks (M\'esz\'aros and Rees, 1997), 
this would be independent of the density of the ambient medium. 
 
\section{Conclusions}

We have detected the optical counterpart to GRB\,980326.
Its temporal decay is well represented by a power law 
with index --2.10, faster than for any previously found GRB afterglow, and a 
constant contribution at $R_{\rm c}$ = 25.5$\pm$ 0.5, which is most likely
caused by an underlying galaxy. 
Fireball models can give an adequate description of this rapid power law
decay of GRB\,980326, although its limited optical spectral information 
makes it hard to distinguish between different models. This emphasizes
the need for multi-colour photometry, even when the optical
counterpart has not yet been found.  

A rapid temporal decay may be a reason 
for the non-detection of low-energy afterglows of bursts that had
X-ray and optical follow-ups.
The occurrence of GRBs in galactic halos, in the merging
neutron star scenario, may be an alternative explanation for the
non-detection of low-energy afterglows.  
To establish the viability of these explanations for the non-detection
of low-energy afterglows, it is of vital importance that more GRB
afterglows are found and this is only possible when low-energy
follow-up begins as soon as possible ($<$1hr) after the initial GRB event. 

\vspace{1cm}
{\bf Acknowledgments} PJG wishes to thank Bram Achterberg for useful
discussions. TJG is supported through a grant from NFRA under contract
781.76.011.  RAMJ is supported by a Royal Society URF grant.  CK
 acknowledges support from NASA grant NAG 5-2560.    

\references

Achterberg, A. and Gallant, Y., 1998, MNRAS, in preparation\\
Bell, A.R., 1978, MNRAS 182, 147\\
Blandford, R.P. and Ostriker, J.P., 1978, ApJ 221, L29\\ 
Briggs, M., et al., 1998, IAU Circ No. 6856\\
Castro-Tirado, A., et al., 1997, IAU Circ No. 6598 \\
Castro-Tirado, A., et al., 1998a, Science 279, 1011\\
Castro-Tirado, A., et al., 1998b, IAU Circ No. 6848\\ 
Celidonio, G., et al., 1998, IAU Circ No. 6851\\
Costa, E. et al., 1997, Nature 387, 783\\
Diercks, A., et al., 1998, ApJL, {\it submitted}, astro-ph/9803305\\
Djorgovski, G., et al., 1998, GCN message\footnote{at 
http://gcn.gsfc.nasa.gov/gcn} No. 41\\
Feroci, M., et al., 1997, Proc. SPIE 3114, 186\\
Feroci, M., et al., 1998, A\&A 332, L29\\
Frontera, F., et al., 1997, A\&AS 122, 357\\
Fruchter, A., et al., 1998, in {\it the 4th Huntsville Symposium on Gamma-Ray
Bursts}, eds. Meegan, Preece and Koshut, 
AIP Conference Proceedings 428, in press.\\
Galama, T.J. et al., 1997, Nature 387, 479\\
Galama, T.J. et al., 1998a, in {\it the 4th Huntsville Symposium on Gamma-Ray
Bursts}, eds. Meegan, Preece and Koshut,  
AIP Conference Proceedings 428, in press, astro-ph/9712322.\\
Galama, T.J. et al., 1998b, ApJ 497, L13\\
Galama, T.J., et al., 1998c, ApJL, {\it in press}, astro-ph/9804190\\
Galama, T.J., et al., 1998d, ApJL, {\it in press}, astro-ph/9804191\\
Garcia, M., et al., 1998, ApJL, {\it in press}\\
Gorosabel, J., et al., 1998, A\&A, {\it submitted}\\
Groot, P.J. et al., 1998a, ApJ 493, L27\\
Groot, P.J. et al., 1998b, IAU Circ No. 6852\\
Grossan, B., et al., 1998, GCN message No. 35\\
Halpern, J.P. et al. 1998, Nature, 393, 41\\
Hurley, K., et al., 1998, GCN message No. 53\\
In 't Zand, J.J.M., et al., 1998, {\it submitted}\\
Jager, R., et al., 1997, A\&AS 125, 557\\
Kulkarni, S.R. et al., 1998, Nature 393, 35\\
Landolt, A.U., 1992, AJ 104, 340 \\
Mallozzi, R.S., Pendleton, G.N., Paciesas, W.S., Preece R.D. and
Briggs, M.S., 1998, in {\it the 4th Huntsville Symposium on Gamma-Ray
Bursts}, eds. Meegan, Preece and Koshut, 
AIP Conference Proceedings 428, {\it in press}\\
Marshall, F. and Takeshima, T., 1998, GCN message No. 58\\ 
Meegan, C.A., et al., 1996, ApJS 106, 65\\
M\'esz\'aros, P. and Rees, M.J., 1997, ApJ 476, 232\\
Metzger, M.R. et al., 1997, Nature 387, 879\\
Nicastro, L., 1998, A\&A, {\it submitted}\\
Paczy\'nski, B., 1998, ApJ. 494, L45\\
Pedersen, H., et al., 1998, ApJ 496, 311\\
Pian, E., et al., 1998, ApJ 493, L103\\
Piro, L., Scarsi, L., and Butler, R.C., 1995, Proc. SPIE 2517, 169\\
Quenby, J.J. and Lieu, R., 1989, Nature 342, 654\\
Rhoads, J.E., 1998, in {\it the 4th Huntsville Symposium on Gamma-Ray
Bursts}, eds. Meegan, Preece and Koshut, AIP Conference Proceedings
428, {\it in press}, astro-ph/9712042\\  
Rybicki, G.B. and Lightman, A.P., 1979, Radiative Processes in
Astrophysics, Wiley, New York, Chapter 6\\
Sahu, K.C. et al., 1997, Nature 387, 476\\
Sari, R., Piran, T., and Narayan, R., 1998, ApJ 497, L17\\ 
Sokolov, V.V., et al., 1998, A\&A, {\it in press}, astro-ph/9802341\\
Van Paradijs, J. et al., 1997, Nature 368, 686\\
Wijers, R.A.M., Rees, M.J. and M\'esz\'aros, P., 1997, MNRAS 288, L51\\
Yoshida, A., et al., 1998, in {\it the 4th Huntsville Symposium on Gamma-Ray
Bursts}, eds. Meegan, Preece and Koshut, AIP Conference Proceedings
428, {\it in press}\\  

\small
\begin{table*}
\caption[]{Log of observations of GRB\,980326, supplemented with published 
observations of the Keck II and KPNO 4-m telescopes.}
\begin{minipage}{15cm}
\begin{tabular}{lllll}
\hline
Date (UT)	& Telescope & Integration time (s) & Magnitude
OT & Reference\\
Mar. 27.31 	& Keck II 	& 	& $R_{\rm c}$=21.19 $\pm$ 0.1
& GCN \#33 \\
Mar. 27.401 	& AAT     	& 240   & $R_{\rm c}$=21.98 $\pm$ 0.16  	&\\
Mar. 27.437 	& AAT     	& 240	& $R_{\rm c}$=22.18 $\pm$ 0.16  	& \\
Mar. 27.84	& BO 1.5m 	& 3600  & $R_{\rm c}>$21.85
& GCN \# 42 \\
Mar. 27.852      & CAHA         & 3300  & $R_{\rm c}>$22.0 \\
Mar. 28.016 	& ESO NTT 	& 1200  & $R_{\rm c}$=23.66 $\pm$ 0.12 	&\\
Mar. 28.017     & ESO 1.5Dan    & 2700  & $R_{\rm c}$=23.43 $\pm$ 0.25    &\\ 
Mar. 28.045 	& CTIO 4m 	& 600	& $R_{\rm c}$=23.50 $\pm$ 0.12 	&\\
Mar. 28.120 	& FLW 1.2m 	& 3600	& $R_{\rm c}>$22.5 		&\\
Mar. 28.178 	& ESO NTT 	& 1200	& $R_{\rm c}$=23.60 $\pm$ 0.12    &\\
Mar. 28.25 	& Keck II 	& 	& $R_{\rm c}$=23.69 $\pm$ 0.1
&GCN \# 32\\
Mar. 29.09      & CTIO 4m       & 3120  & $B$=25.03$\pm$0.33    &\\
Mar. 29.035     & ESO NTT       & 1800  & $I_{\rm c}>$22.4              & \\
Mar. 29.008     & ESO NTT       & 1800  & $V>$24.2              &\\ 
Mar. 29.424 	& AAT     	& 480	& $R_{\rm c}>$23.0		&\\
Mar. 30.078     & ESO NTT       & 5400  & $R_{\rm c}$=\gpm{24.88}{0.32}{0.26}     &\\
Mar. 30.2  	& Keck II 	& 	& $R_{\rm c}$=25.03 $\pm$ 0.15
&GCN \#35\\ 
Mar. 31.082     & ESO NTT	& 5400  & $R_{\rm c}$=\gpm{25.20}{0.23}{0.20}    &\\
Apr. 1.080      & ESO NTT       & 5400  & $R_{\rm c}>$24.9     & \\
Apr. 7.15 	& KPNO 4m	& 3300  & $R_{\rm c}>$24.4&\\
Apr. 17.3 	& Keck II	&       & $R_{\rm c}$=25.5$\pm$0.5 &
GCN \#57
\end{tabular}
\end{minipage}
\end{table*}

\begin{table}
\begin{minipage}{15cm}
\caption[]{The magnitudes of the four comparison stars used\footnote{
Photometric calibration of our observations was
performed using Landolt (1992) standard fields SA98 and Rubin 149
($R_{\rm c}$-band, taken at the AAT at Mar. 27.4 UT), and PG1047+003 ($B$,
$V$ and $I_{\rm c}$-band, taken at ESO at Mar. 30.05 UT).}}
\begin{tabular}{lllll}
\hline
Star no.
 & $B$	& $V$ 	& $R_{\rm c}$	& $I_{\rm c}$	\\\hline
1&20.05$\pm$0.10  & 19.17$\pm$0.07     &18.51$\pm$0.03     &18.11$\pm$0.02\\
2&-               & 23.04$\pm$0.15     &21.85$\pm$0.10     &20.74$\pm$0.05\\
3&21.08$\pm$0.10  & 20.76$\pm$0.05     &20.40$\pm$0.05     &20.00$\pm$0.02\\
4&20.73$\pm$0.10  & 20.22$\pm$0.05     &19.78$\pm$0.03     &19.53$\pm$0.02\\
\end{tabular}
\end{minipage}
\end{table}

\end{document}